\documentclass[a4,10pt]{article}%edp-jp4}
 
\usepackage{pst-node,amssymb,hyperref,cite,fullpage}%,marvosym,fancybox}
 
\newcommand{\ket}[1]{\ensuremath{|#1\rangle}}

\newcommand{\NZ}{{\field{N}}_0}
\newcommand{\field}[1]{\mathbb{#1}}

\newtheorem{definition}{Definition}
\newtheorem{conjecture}{Conjecture}

\begin{document}
\title{Deutsch's Universal %and Programmability of 
Quantum %Computers (Revisited)}
Turing Machine (Revisited)}
% For Durban quantum conference, July 2006
% Revised for proceedings, September-November 2006
\author{Willem Fouch\'e (Department of Decision Sciences)\\
Johannes Heidema (Department of Mathematical Sciences)\\
Glyn Jones (Department of Physics)\\
and Petrus H.  Potgieter (Department of Decision Sciences, University of South Africa)\\
php\,@\,member.ams.org (corresponding author)\\
PO Box 392, Unisa 0003, Republic of South Africa.}
\maketitle
 
\begin{abstract}
Deutsch, Feynman, and Manin viewed quantum computing as a kind of universal physical simulation procedure. Much of the writing about %quantum logic circuits and 
quantum Turing machines has shown how these machines can simulate an arbitrary unitary transformation on a finite number of qubits. This interesting problem has been addressed most famously in a paper by Deutsch, and later by Bernstein and Vazirani. 
% as well as Kitaev and Solovay. The quantum logic circuit model, also initiated by Deutsch, has been more prominent in the research literature than quantum Turing machines. 
Quantum Turing machines form a class closely related to deterministic and probabilistic Turing machines and one might hope to find a universal machine in this class. A universal machine is the basis of a notion of programmability. The extent to which universality has in fact been established by the pioneers in the field is examined and a key notion in theoretical computer science (universality) 
is scrutinised.
%in the study of quantum computing is scrutinised by distinguishing various connotations and concomitant results and problems.
%\keywords{universal quantum computer, programmability, universality, quantum Turing machines, quantum gates}
%%
%% Volgende sin ingevoeg.
%%
In a forthcoming paper, the authors will also consider universality in the quantum gate model.
\end{abstract}
 
\section{Introduction}
 
%By the beginning of the twentieth century mathematicians had become quite interested in establishing a formal model of computability. 
In 1936 Alan Turing described an abstract device, now called a \textit{Turing machine}, which follows a simple, finite set of rules in a predictable fashion to transform finite strings (input) into finite strings (output, where defined). 
The Turing machine (TM) can be imagined to be a small device running on a two-way infinite tape with discrete cells, each cell containing only the symbol \textbf{0} or \textbf{1} or a blank. %The TM 
It has a finite set of possible internal states and a head that can read the contents of the cell of the tape immediately under it. The head may also, at each step, write a symbol to the cell over which it finds itself. There are two special internal states: an \textit{initial state} $q_0$ and a \textit{halting state} $q_H$. 
A TM has a finite list of instructions, or \textit{transition rules}, describing its operation. There is at most one transition rule for each combination of cell content (under the head) and internal state. If the internal state is $q_{i}$ and the head is over a cell with content $S_{j}$ then the machine looks for a rule corresponding to $(q_i, S_j)$. If no rule is found, the machine enters the halting state immediately. If a rule corresponding to $(q_i, S_j)$ is found, it will tell the machine what to write to the cell under the head, whether to move left or right and which internal state to enter. There is no transition rule corresponding to the halting state. Sometimes we refer to the entire collection of individual rules for all the different $(q_i, S_j)$ as \textit{the transition rule} of the machine. 
A \textit{computation} consists of starting the TM with the head over the first non-blank cell 
%(which we may label position 0 on the tape) 
from the left of the tape (it is assumed that there is nothing but some finite \textit{input} on the tape) and the machine in internal state $q_0$. Now the transition rules are simply applied until the machine enters the halting state $q_H$, at which point the content of the tape will be the \textit{output} of the computation. 
%If, for some input, the machine never halts then the output corresponding to that input is simply undefined. 
It is clear how every TM defines a (possibly, partial) function $f:\NZ\rightarrow\NZ$  from the set of counting numbers to itself.
 
Turing machines are the canonical models of computing devices. No deterministic device, operating by finite (but possibly unbounded) means has been shown to be able to compute functions not computable by a Turing machine. 
%In fact, one may view one's desktop computer as a Turing machine with a \textit{finite} tape. 
A \textit{probabilistic Turing machine} (PTM) is identical to an ordinary Turing machine except for the fact that at each machine configuration $\left(q_i,S_j\right)$ there is a finite set of transition rules (each with an associated probability) that apply and that a random choice determines which rule to apply. We fix some threshold probability greater than even odds (say, 75\%) and say that a specific PTM computes $f(x)$ on input $x$ if and only if it halts with $f(x)$ as output with probability greater than  75\%. 
 
\section{Quantum Turing Machines (QTMs)}
\label{secqtm}
%\marginpar{Johannes het Petrus baie gehelp hier om die terminologie, t.w. "steps" edm. beter te omskryf en ondubbelsinnig te gebruik.}
 
A natural model for quantum computation is based on the classical Turing machine. The \textit{quantum Turing machine} (QTM) was first 
%\footnote{Paul Benioff had related a similar idea somewhat earlier \cite{Ben80a} but primarily in connection with presenting a possible physical basis for reversible computing.} 
described by David Deutsch \cite{Deu85a}.  The basic idea is quite simple, a QTM being roughly a probabilistic Turing machine (PTM) with complex transition amplitudes 
%(the squared moduli of which add up to one at each application) 
instead of real probabilities. 
 
\subsection{Operation of a QTM}
 
The QTM is related to the classical deterministic TM in much the same way as the PTM is. In the following the \textit{classical machine} is a machine with a two-way infinite tape, starting over position 0 on the tape as described above. A corresponding quantum Turing machine (QTM) might work as follows (based on the Deutsch description \cite{Deu85a}, Ozawa \cite{Oza97a}, Bernstein and Vazirani \cite{BV97a}).
 
\begin{enumerate}
\item The quantum state space of the machine is spanned by a basis consisting of states 
$$\ket{h}\ket{q_C}\ket{x_C}\ket{T_C}$$ 
where $h\in\{0,1\}$ and $(q_C,x_C,T_C)$ is a configuration of the corresponding classical machine, where $x_C$ denotes the position of the head, $q_C$ the internal state of the machine and $T_C$ the non-blank content of the tape. $T_C$ should include an indication of the absolute position of the content on the tape.
%(otherwise the description would be incomplete). 
\item Special initial and terminal internal states have been identified.
% (corresponding to the initial state and halting state of the classical machine).
\item The single  transition rule is now a unitary operator which, in each step, maps each basic $\ket{h}\ket{q}\ket{x}\ket{T}$ to a superposition of finitely many  $\ket{h'}\ket{q'}\ket{x'}\ket{T'}$, where 
\begin{enumerate}
\item $T'$ and $T$ differ at most in position $x$;
\item $|x'-x|\leq 1$;
%(depending on whether the corresponding classical machine moves one position to the left, to the right, or not at all);
\item $h'=1$ whenever $q'$ is the halting state of the classical machine; and
\item $T'=T$, $q'=q$ and $h'=h$ whenever $h=1$.
\end{enumerate} 
Note that the transitional rule (``program'') will have a finite specification only if the transition amplitudes in the superposition of the $\ket{h'}\ket{q'}\ket{x'}\ket{T'}$ are all \textit{computable} complex numbers, which we will of course assume to be the case throughout. The transition rule can also, obviously, be extended (linearly) to finite superpositions of $\ket{h}\ket{q}\ket{x}\ket{T}$.
\item The machine is started with a finite superposition of inputs in the initial state. Because of the form that the transition rule is  allowed to take %(and the fact that there are only finitely many internal machine states) 
the machine will be in the superposition of  only \textit{finitely many} basic states $\ket{h}\ket{q}\ket{x}\ket{T}$ at any step during the entire run 
%\footnote{A more hazy concept than for classical Turing machines, as a QTM only really stops when one has observed the halting qubit and the content of the tape, so one may think of the transition rule being applied \textit{ad infinitum}, step-by-step, unless the operator (physically, classically and externally) stops the machine.} 
of the computation.
\end{enumerate} 
Without loss of generality everything can be assumed to be coded in binary so that each position on the tape will 
correspond to a single qubit (quantum bit).  A unit of quantum information, the qubit is a two level quantum mechanical system, whose state is described by a linear superposition of two basis quantum states, often labelled \ket{0} and \ket{1}.
The actual (quantum) state space of the machine will be a direct sum of $n$-qubit spaces (where $n$ is an indication of how much tape has been used, each $n$-qubit space being the $n$-fold tensor of the single qubit space). 
%The direct sum is, however, not a complete inner-product space and therefore---by the postulates of quantum mechanics---not a valid state space. However, the underlying Hilbert space can be taken to be the completion of the direct sum and a unitary operator $U$ on the direct sum (see \cite{BV97a}) can be extended to a unitary operator $\widehat{U}$ on the Hilbert space. This completed space and operator will correspond to the physical system associated with the QTM, thereby taking care of the \textit{physicality} of the QTM, as embedded in its Hilbert space completion. 
%The operator $U$ describes only one step in the operation of the machine.
 
\subsection{Time evolution of the QTM and halting}
 
If $U$ is the operator that describes one application of the transition rule (i.e. one step in the operation) of the machine, then the evolution of an  unobserved machine (where not even the halt bit is measured) for $n$ steps is simply described by $V=U^n$. If the first measurement occurs after $n_1$ steps, and the measurement is described by an operator $J_1$ then the evolution of the machine for the first $n_1+j$ steps is described by 
%Die indruk van die uitdrukking $U^jJ_1U^{n_1}$ in die lyn i.p.v. "display mode" ongelukkig nodig
$U^jJ_1U^{n_1}$, which is in general no longer unitary since the operator $J_1$ is a measurement (always in the computational basis). It is important to note that the machine evolves unitarily only when no measurement takes place at all. 
 
The output of the machine is on the tape as a superposition of basis states and should be read off after having measured the content of the halt bit and finding it in the state 1. The operator may at any time measure the halt bit\footnote{
The halt \textit{qubit}, of course, until we measure it.} in order to decide whether to read the tape content (and collapse the state of the machine to one of the basis states). The halt bit is intended to give the operator of the machine an indication of when an output may be read off from the tape (and by observation collapsing the system to an eigenstate) without interfering excessively with the computation. It seems that Deutsch's original idea was that there would be no entanglement at all between the halt bit and the rest of the machine, but this cannot be guaranteed. 
The \textit{output} of a QTM for some specific input $x$ (which may be a superposition of classical inputs) is a probability distribution $P_x$ over all possible contents of the tape at the time of observing the halt bit to have been activated. 
%Note that the observation of an activated halting bit may in itself be a random event, but it has been argued by Ozawa and others, that $P_x$ does not depend on the random \textit{observation} events. 

\section{Universality and programmability in the machine model}
 
The notion of a \textit{universal} computing device in a specific class is crucial for the development of a complexity theory and---more basically---establishes the notion of programmability. 
%Naturally, we will start the discussion by reviewing the well-established notions of universality in classical deterministic and probabilistic computing before moving on to examine the concept of a ``universal QTM'' introduced by Deutsch.
 
\subsection{Classical universality and programmability}
 
Consider a general countable class of machines, say \textit{Manchester machines } (MMs), that compute partial functions, i.e. functions that are not necessarily defined for all inputs (since the machine might not halt, for example, as in the case of a Turing machine). 
Since there are only countably many machine descriptions, let us assume that each Manchester\footnote{Alan Turing worked on building and programming one of the first electronic computers in the city of Manchester after the Second World War.} 
machine is fully described by a natural number. It should be possible to recover the full description of the machine's functioning from the natural number in an effective way, so it should not simply be any enumeration of the countable set. Let $\Phi_n$ denote the partial function computed by machine $n$ and fix an MM-computable bijective function %\footnote{%
%It will strike the attentive reader that $h$ is the first (and last) function of two variables to appear here but that we have implicitly considered Manchester machines with one natural number input only. This is illogical, but the problem can be fixed in a well-established way. Suffice it to say that one should be able to consider $h$ MM-computable in an obvious and logical way. One only needs \textit{one} such function $h$ and we will therefore not elaborate here.
%}
$h : \NZ \times \NZ \rightarrow \NZ$,
assuming such a function exists\footnote{$h:(x,y)\mapsto2^x3^y$ would suffice, for example.}.
%% Voetnota verkort
%\footnote{If it does not, the class of machines would really be very poor. It would not make a big difference if we took, for example, $h:(x,y)\mapsto2^x3^y$ instead of an onto function but the convention that $h$ be onto is harmless and convenient.}
\begin{definition}
\label{defone}
  If there exists a number $N$ such that
  \[ \Phi_N \left( h ( n, m ) \right) = \Phi_n ( m ) \]
  which means that the functions are either equal and both defined or both
  undefined, for all $n$ and $m$, then the machine described by $N$ is called a
  \textit{Universal Manchester Machine} (UMM).
\end{definition}
%For the class of Turing machines, a universal machine (in fact, infinitely many) exists. 
Programmability is firmly linked to the concept of universality and is, of course, a necessary condition for universality.  Is it a sufficient condition? 
A particular Turing machine is usually thought of as dedicated to a particular task, defined by a set of quintuples describing the operations to be carried out in  sequence.  Every Turing machine has thus a finite description (of internal states, tape entries and operation rules---which are unbounded but finitely many) which could be used as input to another Turing machine. 
A universal Turing machine, (of which there are infinitely many), can simulate all the Turing machines, and is thus  programmable for the entire class of Turing machines.  If a machine is programmable for any device in its class, then it is universal. Not all programmable computer devices are universal in any  sense. In fact, one could use the term programmable to describe any device taking input of the form $\langle p,x\rangle$ where $p$ is the ``program'' and $x$ the ``data'' and where the action of the machine on $x$ depends on $p$.  Such machines are universal (for their class) only if they can---through the suitable choice of $p$---mimic the operation of any other machine in the class.
 
\subsection{Probabilistic Turing machine universality}

Since halting is a probabilistic notion for a QTM, the notion of universality for quantum devices should be akin to that for probabilistic machines. For probabilistic machines, however, Definition \ref{defone} does not directly apply and it is necessary to generalise it as follows.
\begin{definition}
\label{deftwo}
 If there exists a number $N$ such that
 \[ \Phi_N \left( h ( n, m ) \right) = \Phi_n ( m ) \]
 which means that the functions are either equal and both defined or both
 undefined \textbf{(if deterministic) and if not deterministic then the values have the same distribution}, for all $n$ and $m$, then the machine described by $N$ is called a \textit{Universal Manchester Machine} (UMM).
\end{definition}
In the case, for example, of deterministic Turing machines (which are a strict subset of the probabilistic machines) the two notions of universality coincide, of course. The main aim of this section is to discuss this (second) notion of universality for quantum Turing machines (QTMs). 
One can easily show, incidentally, that every function $f$ which can be computed in this sense by a PTM, is also computable by some ordinary TM in the usual sense. Nevertheless PTMs have always been of interest because probabilistic algorithms can often be found that are quite fast by comparison to the best known classical procedure. The class of PTMs is often defined by restricting the probabilities to $\frac{1}{2}$ or $1$ only. In this case the class can also be obtained by taking the ordinary TMs and adding a special write instruction to write one random bit to the tape. 
%The PTMs are often described, in this model, as ``TMs with access to a fair coin toss''. It is easy to see how a universal machine might be described in this class: it would simply be a universal TM equipped with the random output instruction. Such a \textit{universal PTM} (UPTM) could obviously simulate any other ``coin toss'' PTM perfectly, by which is meant that the output of the UPTM would have exactly the same distribution as the output of a PTM for which it is executing a program. 
 
Now, which PTMs should our UPTM be able to simulate exactly? Well, since each PTM should have a finite description, the UPTM need only be able to simulate a countable collection of PTMs. Let us restrict the set of PTMs to those with  \textit{computable} transition probabilities. Each such machine is fully described by the finite set of transition rules and programs for computing each of the associated probabilities. This description is finite---thanks to the restriction of the probabilities to computable numbers. 
Since there is no reasonable way of giving a finite description of PTMs with non-computable transition probabilities, apart from the usual paradoxical definitions  of the type ``one more than the largest number which can be described in thirteen words'', this concludes the discussion for PTMs. 
Introducing arbitrary real transition probabilities makes no sense as it would immediately make any subset of the natural numbers decidable by a probabilistic machine.

\subsection{A universal QTM?}
 
Deutsch introduced a ``universal quantum computer'' (uQC, where u has not been capitalised in order to emphasise the difference between this universality concept and the preceding) in \cite{Deu85a}. The Deutsch uQC is in effect a QTM as in Section \ref{secqtm}, based on a classical UTM with some additional (8 in \cite{Deu85a}) operations that allow any unitary transformation on one qubit to be approximated arbitrarily closely. 
Deutsch showed in the paper that for any given $L$, $\varepsilon>0$ and quantum device $U$ operating on $L$ qubits,  there exists a program $p_L$ for the uQC that (with input $p_L$ followed by any finite superposition of $L$-qubit basic states) approximates the operation of $U$ on the finite superposition of $L$-qubit basic states with accuracy at least $\varepsilon$ (in the inner-product norm). 
% Given an arbitrary QTM the Deutsch scheme does not give a single program but rather an infinite sequence (more about this below) of programs for the uQC! 
This is not the same kind of universality that we have for probabilistic and for deterministic Turing machines and even the concatenation scheme used by Deutsch has been questioned (for example, by Shi  \cite{Shi2002}).
 
Now, if we consider the earlier (second) definition of universality, then there can be no universal machine for the simple reason that in Deutsch's scheme there are uncountably many (transition rules for) QTMs.  For broadly the same reasons as outlined above for PTMs, we shall restrict ourselves henceforth to QTMs with computable transition amplitudes, i.e. transition amplitudes for which both the real and imaginary parts are computable numbers. 
We now fix a scheme for encoding the QTMs and associate any machine $M$ with the smallest 
%\footnote{%
%Two distinct natural numbers may, of course, encode physically identical machines.} 
natural number that encodes it. 
Note that we say that a QTM outputs $y$ with probability $p$ if the probability of \textit{ever} observing the machine to be in the halt state with the tape in state \ket{y} is $p$. 
Does a universal machine for the (restricted) class of QTMs in the sense of Definition \ref{deftwo} exist?
 
Deutsch provided the rather incomplete solution mentioned above. Bernstein and Vazirani \cite{BV97a} have given another partial solution. They showed that there exists a quantum Turing machine $\mathcal U$ (they actually wrote $\mathcal M$) such that  %(\cite{BV97a})
\begin{quote}
``for any well-formed\footnote{Meaning that the time evolution operator is unitary.} 
QTM $M$, any $\varepsilon>0$, and any 
$T$, $\mathcal U$ can simulate $M$ with accuracy $\varepsilon$ for $T$ steps with slowdown polynomial in $T$ and $\frac{1}{\varepsilon}$.''
\end{quote}
The slowdown and the program for $\mathcal U$ \emph{both depend here on the length of the input}. The full Bernstein-Vazirani result could be summarised by the statement that
\begin{quote}
%% 555 f_{\bar{M}}(T,n,\frac{1}{,\varepsilon}) i.p.v. f_{\bar{M}}(T,\frac{1}{,\varepsilon}) hier onder
there exists a QTM $\mathcal U$ such that  for each QTM $M$ with finite description $\bar{M}$, $n$, $\varepsilon$ and $T$ there is a program ${\mathcal P}(\bar{M},n,\varepsilon,T)$ and a function $f_{\bar{M}}(T,n,\frac{1}{\varepsilon})$  (both recursive in their inputs) such that running $\mathcal U$ on input $\ket{{\mathcal P}(\bar{M},n,\varepsilon,T)}\otimes\ket{x}$ where $|x|=n$ for $f_{\bar{M}}(T,n,\frac{1}{\varepsilon})$ steps results---within accuracy $\varepsilon$---in the same distribution over observable states as running $M$ on input $\ket{x}$ for $T$ steps.
\end{quote} 
The simulation is clearly \emph{only approximate}. What Bernstein and Vazirani mean ``with accuracy $\varepsilon$'' is that if $P$ is the probability distribution over all observable states of $\mathcal U$ after $f_{\bar{M}}(T,n,\frac{1}{\varepsilon})$  steps with the given input and $Q$ is the corresponding probability distribution of $M$ after $T$ steps then $$\frac{1}{2} \sum_x |P(x)-Q(x)| ~\leq~ \varepsilon$$
where the summation is over all possible observable states $x$. Again, approximate simulation is quite different from the universality concept for ordinary and for probabilistic Turing machines (with computable probabilities) as in the latter cases the universal machine's simulation was \textit{exact}. 
Running $\mathcal U$ for exactly  $f_{\bar{M}}(T,n,\frac{1}{\varepsilon})$ steps on any input $\ket{{\mathcal P}(\bar{M},n,\varepsilon,T)}\otimes\ket{x}$ will have simulated the running of $M$ on \ket{x} for $T$ steps. We may not let $\mathcal U$  
run for any more steps as the state of the machine might then drift away from the to-be-simulated state of $M$ after $T$ steps. 
This behaviour is rather different from that of the UTM or UPTM---where there is no need to restrict the number of steps executed! 
%What about the input to the machine? In general, the input to a QTM is allowed to be a (finite) superposition of basis states of the tape but the Bernstein-Vazirani theorem quoted here applies to a single state only. This is not a problem: it is straightforward to see that it also applies to a superposition of $m$ basic states (just replacing $\varepsilon$ by $\frac{1}{m}\varepsilon$).
 
Now, the Bernstein-Vazirani machine $\mathcal U$ immediately suggests the following semi-universal hybrid device (SUHD). The device takes the description $\bar{M}$ of a QTM $M$ as well as $x$ and $\varepsilon$ (which may be taken to be rational) as input. The machine operates as follows.
\begin{quote}
\tt 
\begin{tabbing}
$T$:= 1;\\
$n$:= |$x$|;\\
do \= \\
\> compute $P$ := ${\mathcal P}(\bar{M},n,\frac{\varepsilon}{T},T)$;\\
\> compute $S$ := $f_{\bar{M}}(T,n,\frac{T}{\varepsilon})$;\\
\> run $\mathcal U$ on $\ket{P}\otimes\ket{x}$ for S steps;\\
\> signal that quantum part of device may be observed;\\
\> wait a little;\\
\> reset quantum part of device;\\
\> $T$:=$T$+1;\\
while true;\\
\end{tabbing}
\end{quote}
Note that by replacing $\varepsilon$ by $\frac{\varepsilon}{T}$ we have ensured that by simply letting the SUHD run, we will not only be able to observe the simulated behaviour of $M$ for ever longer times, but also with ever-increasing accuracy. However, the SUHD is still not universal for the class of QTMs in the sense of Definition \ref{defone} or Definition \ref{deftwo}. This is true not only for the very obvious reason that its simulation is only \textit{approximate}, but for the much more fundamental reason that we do not know whether it is a QTM itself!
 
The SUHD is a real hybrid device which consists of a classical Turing-type machine and a quantum part. The SUHD is---in a sense---a robot capable of operating a quantum device (which forms part of itself) and there is no reason to think that such a robot cannot be built. The problem lies therein that the robot only gives a signal when we might observe the quantum part of the device. It cannot know whether we have observed the quantum part or not---otherwise the observer would become part of the device...
 
Now, any quantum device operates reversibly. In the case of the SUHD the step ``reset quantum part of device'' is the part which can be problematic in this regard. If the quantum part was not observed during the step ``wait a little'' then the inverse of the evolution operator of $\mathcal U$ can be used to effect such a reset. But, what if the observer(s) did make an observation of the quantum part during ``wait a little''? Now, the  inverse of the evolution operator of $\mathcal U$ will \textit{not} ``reset quantum part of device''. This is really a serious problem. In an ordinary QTM the evolution of the machine continues even when the halt bit has been observed, but for the SUHD even the observation of the halt bit (which may be in a superposed state, although not necessarily entangled with the rest of the machine) renders the operation of the device non-reversible. 
%This is simply because for the ordinary QTM, the evolution operator can continue after the halt bit has been observed without perturbing the probability distribution that has been defined to be the QTM's \textit{output} (according to Ozawa \cite{Oza97a}) since the observation projects one, in a certain sense, only into a specific ($h=0$ or $h=1$) branch of the computation. 
For the hybrid device %it is not that simple since 
the resetting step requires an undisturbed quantum part. If the quantum part has been disturbed at $T=k$, the operation described above will not be able to correctly reset the quantum part of the device and will not execute the loop faithfully for $T=k+1$. 
%Of course, it is always possible for the operator to be instructed to restart the hybrid device after observation, but then we will be dealing with a new kind of bio-hybrid device and not a universal machine at all. 
%In classical computing this would be the equivalent of the user strictly having to reboot the computer each time after looking at the screen, i.e. there would be no \textit{autonomy} of operation. 
Pure quantum computing devices are prevented by  the No Cloning Theorem from copying initial configurations of substems, which precludes the realisation of such a na\"\i ve hybrid operation by a quantum device.
\begin{conjecture}
The SUHD derived from Bernstein and Vazirani's $\mathcal U$ cannot be made to operate reversibly and is therefore not a QTM.
\end{conjecture}
The immediate consequence of the conjecture is that (as yet) no universal machine has been shown to exist in quantum computing and that the notion of universal programmability has not really been established for quantum computing in the QTM model.

\section{Conclusion}
 
Research into quantum computation over the past 20 years has been very successful in stimulating the development of quantum cryptography (already in industrial application), the study of quantum information and the discovery of novel algorithms for traditionally hard and interesting problems such as prime factorisation. One would be wise, however, to heed the words of Andrew Steane \cite{0034-4885-61-2-002}:
\begin{quote}
``The title quantum computer will remain a misnomer for any experimental device
realised in the next twenty years. It is an abuse of language to call even a pocket calculator
a computer, because the word has come to be reserved for general-purpose machines
which more or less realise Turing's concept of the universal machine. The same ought to
be true for QCs if we do not want to mislead people.''
%However, small quantum information processors may serve useful roles. 
\end{quote} 
This paper has attempted to explain why certain (strong and interesting) results in quantum computation still fall short of establishing universality (and programmability) for quantum 
Turing machines.
%computing. 
At the very least, researchers in the field should attempt to explain how the results of Deutsch, Bernstein and Vazirani, 
%Solovay, Kitaev 
and others can be used or expanded to construct a fully programmable universal quantum Turing machine.
%device. In the worst case, one needs to prove that such a fully universal quantum computer does not exist.

\bibliographystyle{plain}
%\bibliography{kb}

\begin{thebibliography}{1}
 
\bibitem{BV97a} E. Bernstein, U. Vazirani, SIAM J. Comp. {\bf 26}, 1411, (1997).
 
\bibitem{Deu85a} D. Deutsch, Proc. R. Soc. Lond. A {\bf 400}, 97, (1985).
 
\bibitem{Oza97a} M. Ozawa, Phys. Rev. Lett. {\bf 80}, 631, (1997).
 
\bibitem{Shi2002} Yu Shi, Phys. Lett. A {\bf 293}, 277, (2002).
 
\bibitem{0034-4885-61-2-002} A. Steane, Rep. Progr. Phys. {\bf 61}, 117, (1998).
 
\end{thebibliography}

\end{document}